\documentclass[11pt,preprint]{aastex}
\usepackage{placeins}
\usepackage{graphicx}

\begin{document}

\title{AN ANALYTIC MODEL FOR BUOYANCY RESONANCES IN PROTOPLANETARY DISKS }
\shorttitle{BUOYANCY RESONANCES}

\author{Stephen H. Lubow} 
\affil{Space Telescope Science Institute, 3700 San Martin Drive, Baltimore, MD 21218}
\email{lubow@stsci.edu}
\and 
\author{Zhaohuan Zhu\altaffilmark{1}}
\affil{Department of Astrophysical Sciences, Princeton University, Princeton, NJ 08544}
\email{zhzhu@astro.princeton.edu}
\slugcomment{\today}

\altaffiltext{1}{Hubble Fellow}

\shorttitle{BUOYANCY RESONANCES}

\begin{abstract}
Zhu, Stone, and Rafikov (2012) found in 3D shearing box simulations 
a new form of planet-disk interaction that they attributed 
to a vertical buoyancy resonance in the disk.
We describe an analytic linear model for this interaction.
%The model adopts a simplified disk structure in which the vertical
%gravity is constant and downward (upward) above (below) the disk midplane.
We adopt a simplified model involving azimuthal forcing that produces the resonance and
permits an analytic description of its structure.
We derive an analytic expression for the buoyancy torque and show that the vertical torque distribution
agrees well with results of Athena simulations and 
a Fourier method for linear numerical calculations carried out with the same forcing. The buoyancy resonance differs from the classic
Lindblad and corotation resonances in that the resonance lies along tilted planes. Its width depends on damping effects 
and is independent of the gas sound speed. The resonance does not excite
propagating waves. At a given large azimuthal wavenumber $k_y > h^{-1}$ (for disk thickness $h$), 
the buoyancy resonance exerts a torque over a region 
that lies radially closer to the corotation radius  than the Lindblad resonance.
Because the torque is localized to the region of excitation, 
it is potentially subject to the effects of nonlinear saturation.
In addition, the torque can be reduced by the effects of radiative heat transfer between the resonant region
and its surroundings.
For each azimuthal wavenumber, the resonance establishes a large scale density wave pattern in a plane within the disk.
\end{abstract}


\section{INTRODUCTION}

Young planets can experience strong gravitational interactions
with surrounding gas residing in a protoplanetary disk \citep{GT80, LP86, W97, LI11, KN12}.
These interactions can lead to structural changes in a disk through the launching of 
waves that can result in shocks and  gap formation. They can also lead to changes in the orbital
properties of the planet, resulting in their radial migration. Such phenomena are caused by the
 resonant forcing of the gas by the planet. The two types of resonances 
that describe this interaction, the Lindblad and corotational,
have been extensively explored.  Both of these resonances involve planar motions and will
occur in a two-dimensional disk in which  the vertical dynamical effects
(perpendicular to the disk orbital plane) are ignored. For planets on circular orbits, the corotation resonance lies
at the orbit of the planet. The gas response at the corotation resonance is of the form of a trapped
radially evanescent wave with a radial drop-off on a scale  of order the disk thickness $h$.
The other form of resonance, the Lindblad resonance, occurs 
where a forcing frequency component due to the planet matches the epicyclic frequency of the gas.
There are infinitely many of these resonances. In the absence of disk self-gravity, these resonances result in the launching of acoustic waves
that transport energy and angular momentum away from the planet. 
The strongest Lindblad resonances lie close to the planet, but cannot
occur closer than $\sim  h$ radially from the orbit of the planet due to pressure effects.

When the disk vertical structure is taken into account, a richer set of waves can be excited at Lindblad resonances
\citep{LO98, Bate02}.
These waves are of the form of r modes (rotation-dominated), g modes (vertical buoyancy-dominated),  p modes (pressure-dominated), and f modes (fundamental). 
For small or moderate 
azimuthal wavenumbers ($\ll h^{-1}$), the Lindblad response is dominated by an f mode, while
the  r-modes and g-modes are 
less strongly excited. In a vertically isothermal disk that undergoes adiabatic
perturbations, as will be considered in this paper, the fundamental mode consists of two-dimensional planar motions.

By means of 3D shearing box simulations, Zhu et al (2012) have recently shown 
that a new form of planet-disk interaction can occur due to buoyancy resonances.
For this type of resonance, the disk vertical motions play a critical role. 
At such a resonance, a planet forcing frequency component matches the free oscillation frequency
of a vertically displaced fluid element that behaves adiabatically. Like the Lindblad case, there are infinitely
many such resonances, one for each azimuthal wavenumber $k_y$. 
But since the free vertical oscillation frequency varies with  height above
the disk midplane, these resonances do not occur at particular radii as in the Lindblad case. Instead they lie along
tilted planes. These resonances radially extend much closer to the planet than Lindblad resonances. The one-sided 
(inside or outside corotation) torque
that results from the buoyancy resonances was found to be comparable to, but smaller than, the usual Lindblad torque.

The goal of this paper is to explore the analytic properties of buoyancy resonances.
The analytic approach provides a verification of the existence of this resonance and insight into its structure.
We consider the linear response of an isothermal disk that undergoes adiabatic
perturbations to a simplified form of forcing.
As in Zhu et al (2012), we consider the gas to reside in a shearing box, 
as described in Section \ref{beq}.
We consider waves for which the azimuthal wavenumber is of order or greater than $h^{-1}$. These waves play
an important role in determining the total torque caused by a planet. 
To make analytic progress, we consider  forcing that is purely
azimuthal and is independent of radius and height in the disk.
In Section \ref{saf}, we analyze  a disk with constant vertical gravity and find separable solutions
for the linear disk response. We obtain
analytic expressions for the structure of the resonance and the torque distribution.
We then consider in Section \ref{safvg} the case of variable vertical gravity and determine the
linear response numerically by means of a Fourier method in radius.
We then show that the torque density in this case agrees well 
with the torque density obtained by an obvious extension
of the  torque density expression in the constant gravity case.
Section \ref{disc} contains a discussion and Section \ref{sum}
contains the summary.

\section{LINEARIZED SHEARING BOX EQUATIONS \label{beq}}

\subsection{Basic Equations}
We consider gas in a 3D shearing box described by Cartesian coordinates $(x,y,z)$ in a frame
that corotates with the disk at some radius $r$ from the central star and generalize the 2D shearing sheet model \citep[e.g.,][]{GT78}. 
The radial coordinate $x$ is defined such that $x=0$ occurs at radius $r$, and  vertical coordinate $z$ is defined
such that $z=0$ lies at the disk midplane. The disk has a characteristic thickness $h \ll r$.
The unperturbed disk is taken to be isothermal with equation of state $p_0(z) = c^2 \rho_0(z)$ with isothermal
sound speed $c$.
The local angular speed of the disk is $\Omega$ and the unperturbed
disk velocity in the corotating frame is $ 2 A \, x \, \bf{e}_y$, with constant shear rate $2 A$. 

We consider the effects of a single azimuthal Fourier component with azimuthal wavenumber $k_y>0$ 
of the gravitational potential
\begin{equation}
\Psi(x, y,z) = \Phi(x,z) \exp{(i k_y y)}.
\label{eq:Phi}
\end{equation} 
We take $\Phi(x,z)$ to be real and thereby determine the phasing of $\Psi(x, y,z)$.
We determine  the $y$ Fourier components of velocity 
 $(u, v, w)$,  density perturbation $\rho$, and  pressure perturbation  $p$ as functions of $x$ and $z$
 that describe the response to this potential.
The linearized steady state shearing box equations  for the $x, y,$ and $z$ motion, 
mass conservation, and heat for an adiabatic gas are respectively
\begin{eqnarray}
%\partial_t u + 
2 i\,A k_y x \, u-2\Omega v & = &-\partial_x \left( \frac{ p}{\rho_0} + \Phi \right), \label{eq:u}\\
%\partial_t v  +
2 i\,A k_y x \,  v+2B u &= & -i k_y \left( \frac{ p}{\rho_0} + \Phi \right), \label{eq:v}\\
%\partial_t w  + 
2 i\,A k_y x \,  w &=& -g\frac{\rho}{\rho_{0}}
-\frac{\partial_z p}{\rho_{0}} -\partial_z \Phi, \label{eq:w}\\
%\partial_t \rho + 
2 i\,A k_y x \, \rho +w\partial_z \rho_{0}
 &=& - \rho_{0} \left(\partial_x u +  i k_y\, v + \partial_z w \right), \label{eq:rho}\\
%\partial_t \left(\frac{p}{p_{0}}-\gamma \frac{\rho}{\rho_{0}} \right)+
2 i\,A k_y x \left(\frac{p}{p_{0}}-\gamma \frac{\rho}{\rho_{0}} \right) &=&
- w \, \partial_z \ln\left(\frac{p_{0}}{\rho_{0}^{\gamma}}\right) = - \frac{ \gamma w \, N^2}{g}, \label{eq:s}
\end{eqnarray}
where
$B = A + \Omega$ is an Oort constant,
$g$ is the vertical disk gravity that can generally be a function of $z$, and $N$ is the vertical buoyancy frequency for gas 
that can also generally be a function of $z$  
\begin{equation}
N(z) = \sqrt{\frac{\gamma-1}{\gamma}}  \, \frac{g(z)}{c},
\label{eq:N}
\end{equation}
with adiabatic index $\gamma$.

Combining equation (\ref{eq:w}) for vertical motion with  the heat equation (\ref{eq:s}), we have that
the density  perturbation is given by
\begin{equation}
\rho =  - \frac{\gamma N^2 p_{0} \, \partial_z p + \gamma \rho_{0}  p_{0} N^2 \partial_z \Phi +  4 A^2\, k_y^2\, x^2 \rho_0 \, g \, p}{\gamma g \, p_{0} ( N^2 -  4 A^2\, k_y^2\, x^2 ) }.
 \label{rho}
\end{equation}
The denominator on the right-hand side of 
equation (\ref{rho}) vanishes at a buoyancy resonance where the buoyancy frequency matches the forcing frequency
 \begin{equation}
 x_{\rm res}(z) =  \pm \frac{N(z)}{2 A k_y}.
 \label{xres}
\end{equation}
The upper (lower) sign is appropriate for an inner (outer) buoyancy resonance, since $A$ is negative.
Equation (\ref{rho}) then contains  possible singularities at buoyancy resonances.

\subsection{Boundary Conditions \label{bc}}

We describe here the boundary conditions that we generally apply. They are similar
to those used in \cite{Zhu12}.
For an inner (outer) buoyancy resonance, we take the outer (inner) $x$ boundary to be located at the corotation
radius defined by $x=0.$
Near the $x=0$ boundary, quantities are assumed to be point symmetric with respect to $(x,y)$.
 This condition means that
\begin{equation}
Re[p(x,z)\exp{(i k_y y)}] = Re[p(-x,z) \exp{(-i k_y y)}]
\label{bcx}
\end{equation}
near $x=0$.
In its application to the boundary at $x=0$, this relation implies 
$
Im(p(0,z)) = 0.
$
A further application of equation (\ref{bcx})  implies that
$Re(\partial_x p(0,z))  = 0$.
These conditions  
can be compactly written as
\begin{equation}
 Re(\partial_x p(0,z))  + i k_y \, Im(p(0,z))=0.
 \label{xbc}
 \end{equation}

For an inner (outer) buoyancy resonance, following Zhu et al (2012) we take the  inner (outer) $x$ boundary 
condition to be that the radial velocity perturbations vanish, $u=0$. 
By combining equations (\ref{eq:u}) and (\ref{eq:v}),
we obtain a condition on the
pressure perturbation at this boundary that
\begin{equation}
\partial_x p = -\frac{A  \, x \, \rho_0 \, \partial_x \Phi +  \Omega \, \rho_0 \, \Phi  +  \Omega \,p}{A  \, x}.
\label{xobc} 
\end{equation}

Periodic boundary conditions are applied in the $y$ direction. This condition is automatically
handled by our use of Fourier components in $y$. 

In the $z$ direction,  
we apply reflection boundary conditions at the disk midplane. That is,
\begin{equation}
w(x,0) = 0.
\label{zbc}
\end{equation}
This boundary condition implies that there is no mass flux
through the disk midplane from above or below.
Far from the disk midplane (large  $|z|$), the disk pressure perturbations are assumed to vanish,  $p=0$.

\subsection{Torque}

To determine the torque due to a buoyancy resonance, we determine $\rho$ near $x_{\rm res}$.
We assume, as we later show, that the numerator on the right-hand side of equation (\ref{rho})
does not vanish at a buoyancy resonance.
We expand that equation about $x=x_{\rm res}$ and obtain to lowest order that
\begin{equation}
\rho =  \pm N \frac{\gamma p_{0} \, \partial_z p  + \gamma \rho_{0}\,p_{0} \,  \partial_z \Phi  + \rho_{0} \, g \, p}{4 \gamma g \, p_{0}  A k_y  x'  } ,
 \label{rhoex}
\end{equation}
where $x'= x - x_{\rm res}$ is the $x$ position relative to the resonance.
To treat the singular behavior of $\rho$ at $x'=0$, we follow the standard procedure of
extending $x'$ to the complex plane and replacing $x'$ by $x' + i \, \l_{\rm d}$  with small damping length $\l_{\rm d}$
\citep[e.g.,][]{MS87}.
%This procedure is equivalent to considering the effects of a weak dissipative force in which $\l_{\rm d}$
%must be real and positive \citep[e.g.,][]{MS87}. For a Keplerian disk, it can be shown that $l_{\rm d} ^{-1}=  1.5 k_y \Omega  \, t_{\rm d}$, where
%$t_{\rm d}$ is a damping timescale associated with the dissipation.
%We take $x'$ into the complex plane, $x' \rightarrow x' + i \epsilon$. 
We then have that 
\begin{equation}
\rho =  \pm N \frac{\gamma p_{0} \, \partial_z p    + \gamma \rho_{0}\,p_{0} \,   \partial_z \Phi+ \rho_{0} \, g \, p}{4 \gamma g \, p_{0}  A k_y } 
\left[ \frac{x'}{x'^2+\l_{\rm d}^2}-\frac{i \, \l_{\rm d}}{x'^2+\l_{\rm d}^2} \right].
\label{rhoc}
\end{equation}
The second term in the brackets represents the "resonant" term that behaves as a Dirac delta function, since
\begin{equation}
\delta(x')= \frac{1}{\pi} \frac{\l_{\rm d}}{x'^2+\l_{\rm d}^2}
\end{equation}
for small $\l_{\rm d}$. 

The  torque density in $z$ inside/outside corotation on the gas due to a particular $y$ Fourier component
of the perturbing potential $\Psi$ is defined by
\begin{eqnarray}
\frac{d T}{dz} &=&  -r \,  \int_{-\pi r}^{\pi r} \int_{x_{\rm i}}^{x_{\rm o}} Re[\rho(x,z) \exp{(i \, k_y y)}] \, Re[\partial_y \Psi(x,y,z)] \, dx\, dy\\
 &=&  -\pi r^2 k_y  \int_{x_{\rm i}}^{x_{\rm o}}  Im(\rho) \Phi \, dx,
\label{T}
\end{eqnarray}
where we used the fact that $\Phi$ is a real quantity in equation (\ref{eq:Phi}). The $x$ integration limits apply
either inside or outside corotation where $x_{\rm i}$ is the $x$ inner boundary location
and $x_{\rm o}$ is the $x$ outer boundary location. For the case inside (outside) corotation,
$x_{\rm o} = 0$  ($x_{\rm i} = 0$).  
We then obtain
\begin{equation}
\frac{d T}{dz} = \pm  \frac{\pi^2 r^2  \Phi  N}{4 \gamma g \, p_{0}  A}   \left(\gamma p_{0} \, Re(\partial_z p)    + \gamma \rho_{0}  \, p_{0} \, \partial_z \Phi+ \rho_{0} \, g \, Re(p) \right),
\label{Ti}
\end{equation}
where all $x$-dependent quantities are evaluated at the resonance where $x=x_{\rm res}$. 
To evaluate the torque, we determine the pressure perturbation $p(x,z)$ near the resonance.

\section{MODEL WITH CONSTANT VERTICAL GRAVITY AND SIMPLE AZIMUTHAL FORCING \label{saf} }

\subsection{Torque Derivation \label{TDcg}}
In a standard thin disk, the buoyancy frequency $N$ varies nearly linearly with $z$, due to the change in vertical gravity 
with $z$ (see equation (\ref{eq:N})).
The resonance condition (\ref{xres}) is then satisfied along tilted planes.
We simplify the geometry of the resonances by applying an approximate model for the vertical gravity.
Above (below) the disk midplane, the vertical gravity is taken to be constant and downward (upward). In the analysis below,
we consider the dynamics above the disk midplane where the downward vertical gravity is denoted by the constant $g>0$.
%In this case, the buoyancy frequency is a postive (negative) constant  above (below)
%the disk midplane. In this case,  t
The resonance location
simplifies to a vertical plane described by a particular value of  $x$ for each $k_y$. 
The constant vertical gravity model permits the dynamical equations to be separable in space
and facilitates the development of an analytic model.

Vertical hydrostatic balance for constant gravity implies that the unperturbed disk satisfies
\begin{eqnarray}
p_0(z) &=& p_{00} \exp{(-|z|/h)}, \label{p0}\\
\rho_0(z) &=& \rho_{00} \exp{(-|z|/h)}, \label{rho0}
\end{eqnarray}
where 
\begin{equation}
c^2 = g \, h,
\end{equation}
with constants $g$, $h$, $p_{00}$ and $\rho_{00}.$ 
We also take 
\begin{equation}
g = \Omega^2 \, h.
\end{equation}

To make analytic progress, we adopt a simple potential
of the form (\ref{eq:Phi}).
We take the perturbing potential to
be of the form
\begin{equation}
\Psi(x,y,z) = \Phi \exp{(i k_y y)},
\label{eq:Phic}
\end{equation}
where $\Phi$ on the right-hand side is a real constant.
This potential gives rise to forcing that is purely azimuthal
and is constant in $x$ and $z$.  

To investigate the nature of this simplification, we carried out some
nonlinear numerical simulations using the Athena code.
The simulations were of isothermal disks which are subject to various vertical gravities
 and undergo adiabatic perturbations caused by various potentials. 
  We define dimensionless wavenumber and coordinates as
  $K_y = k_y h$,  $X= x/h$ (a different $X$ than defined in Section \ref{ns}),  $Y=y/h$, and $Z=z/h$.
 We express the perturbing potential in terms of the modified Bessel function $K_{0}$.
 These simulations covered cases with \\
 (a) constant vertical gravity $g$ and potential given by equation (\ref{eq:Phic}),
 so that $\Phi(X,Z)=C$, \\
(b) variable vertical gravity and potential given by $\Psi(x,y,z)=C \,K_{0}(k_{y}  x) \,\exp(i k_{y} y),$
so that $\Phi(X,Z)=C \,K_{0}(K_{y}  X)$, and\\
 (c) variable vertical gravity and potential given by  $\Psi(x,y,z)=C \, K_{0}(k_{y}\sqrt{x^2+z^2}) \, 
 \exp(i k_{y}y),$ so that $ \Phi(X,Z)=C \,K_{0}(K_{y}  \sqrt{X^2+Y^2}$).\\
  The disk structure and the  potential perturbation
become more realistic in going from Case (a) to (c). 
 
 Constant $C$ is chosen as 5.8$\times$10$^{-3}\,c^2$. 
 For Case (c), this is equivalent to the $k_{y}$
Fourier component of the potential of a planet with mass $5.8\times10^{-3}\pi M_{*}(h/r)^{3}$ or 0.76 $M_{\oplus}$
with $h/r=0.05$ and $M_{*}=M_{\odot}$.
 This value is sufficiently small
  that there is no gap opening. 
 In Cases (b) and (c), a small smoothing length 5$\times$10$^{-3} h$ has been used.
  A small isotropic viscosity  has been used to resolve the resonance. The viscosity is $\nu=10^{-6} \, \Omega h^2$ 
  (or equivalently $\alpha = 10^{-6}$) for Case (a), while $\nu=10^{-5} \,\Omega h^2$  (or equivalently $\alpha = 10^{-5}$)
  for Cases (b) and (c), respectively.

 The simulated gas lies  above the disk midplane $z>0$  and outside corotation $x>0$.
 The simulation setup was similar to that described in Zhu et al (2012). 
For Case (a), we used 32 vertical grid points per scale height that gives a total torque 
that is nearly independent of the resolution (the error is within 5\%). 
However, the simulation does not resolve the vertical torque structure to very high accuracy. The peak of the
torque density as a function of height $z$ for the $k_{y}h=2\pi$ mode is at $z \sim 0.2 h$. For constant vertical gravity, we find that
this resolution corresponds to about 6 vertical grid points near the peak of the torque density. 
%Furthermore, the required
%resolution in the z direction depends on the $k_{y}$ mode studied. Ideally, we should have a large number
%of grids with each wavelength.
 
 The boundary
 conditions follow those described in Section \ref{bc}. The simulation domain and resolution (in $x,y,z$ order)  for these
 three simulations are respectively: (a) $(0, 0.3 h) \times (-0.5 h, 0.5 h) \times (0, 5 h)$ with the resolution of
 $252\times128\times640$, (b) $(0, 0.3 h) \times (-0.5 h, 0.5 h) \times (0, 5 h)$ with the resolution of
 $126\times64\times320$, and (c) $(0, 2 h) \times (-0.5 h, 0.5 h) \times (0, 5 h)$ with the resolution of
 $256\times64\times320$.
 In these simulations, 
 we take $k_y h = 2\pi$ and $\gamma = 5/3$. 
 
\begin{figure}
\epsscale{1.0} \plotone{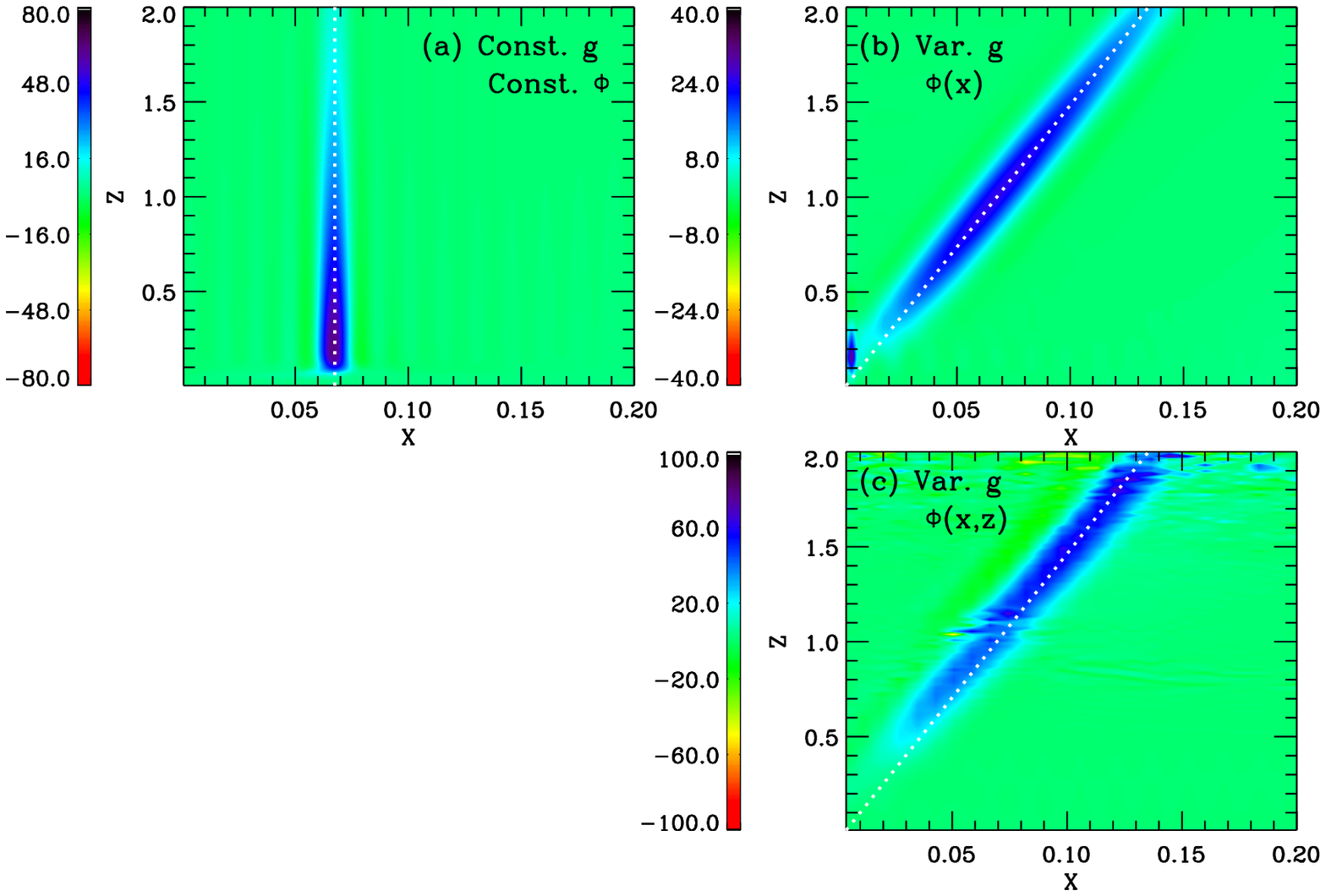} \caption{
$y$-integrated torque density in the $x-z$ plane
obtained from nonlinear numerical Athena simulations
for $k_y h = 2 \pi$ and $\gamma = 5/3$ for the three cases described in
Section \ref{TDcg}. The torque density $\partial_{X} \partial_{Z} T(X,Z)$ is scaled by $r^2 h \rho_{00} \Phi^2(X, Z) /c^2$ for $X=x/h$ (a different $X$ from that in Fig.~\ref{fig:px}) and $Z=z/h$ (see equation (\ref{dTdXdZ})).
Panel (a) is the case of constant gravity and simplified potential given by equation (\ref{eq:Phic})  that is independent of $x$ and $z$.
Panel (b) is the case of vertically varying gravity and simplified potential that is independent of $z$. 
Panel (c) is the case of vertically varying gravity and the potential is the $k_y$ Fourier component
of the planet potential. 
The artifact 
within $x<0.005 h$ of Panel (b) is due to the singularity in the cylindrical potential
and the small smoothing length  $5\times$10$^{-3} h.$  
The potentials are chosen to be weak enough 
to prevent gap opening.
In all cases, the torque is confined to a thin region near the plane of the buoyancy resonance location (dotted line)
defined by equation (\ref{xres}).  }\label{fig:tdens}
\end{figure}

 The simulations were conducted in
 dimensionless units with the 
 dimensionless gas density $\rho_{\rm s}$ expressed in units
  of the unperturbed midplane density $\rho_{00}$, the dimensionless coordinates expressed
  in units of length $h$,
  and dimensionless time in units of $\Omega^{-1}$. Accordingly,
  the dimensionless potential used in the simulations
  is $\Phi_{\rm s} = \Phi/c^2$.
  From the simulation results, we determine the torque per length in the $y$
  direction that we multiply by $2 \pi r$ to obtain the torque density 
  $\partial_{X} \partial_{Z} T$.  We expect the torque to scale with $\Phi^2$,
  since both the density perturbation and $y$ gravitational force depend linearly on $\Phi$.
  In Fig.~\ref{fig:tdens}, we plot the
  dimensionless torque density that is scaled by $\Phi_{\rm s}^2$ and is given by
  \begin{equation}
  \partial_{X} \partial_{Z} T_{\rm s} =  \frac{\pi K_y} {\Phi_{\rm s}(X, Z)} \, \frac{1}{2 Y_{\rm b}}\int_{- Y_{\rm b}}^{ Y_{\rm b}} \rho_{\rm s} (X,Y,Z) \sin{(K_y Y)} \, dY,
  \label{dTdXdZ}
  \end{equation}
  where $\pm Y_{\rm b}$ are the dimensionless locations of the $y$-boundaries of the simulations and $\rho_s$ is the 
  simulated disk density with both expressed in dimensionless units.
  This torque density is converted to the dimensional torque density $\partial_{X} \partial_{Z} T$ by multiplying by
  $r^2 h \rho_{00}  \Phi^2(X,Z)/c^2$.
  
Fig.~\ref{fig:tdens} provides some motivation for the gravity and potential simplifications.
In all three Cases (a), (b), and (c), plotted in the respective panels after 7 orbits, the torque is confined to a narrow region
centered on the buoyancy resonance.
When scaled by $r^2 h \rho_{00} \Phi^2(X,Z)/c^2$,
the torque densities have somewhat similar structures. 
The vertical orientation of the resonance
in the case of constant gravity permits an analysis through separation of variables.
%Although the simple potential given by equation (\ref{eq:Phic})
%is quite different from the corresponding potential for a point mass, it provides
%the same qualitative behavior at a buoyancy resonance.  

We concentrate on the case that $k_y$ is large, $k_y \ga h^{-1}$.
We are interested in this regime because the total torque is dominated by contributions that occur
at large $k_y$.
The  gas at resonance has sonic or subsonic unperturbed gas speeds at the resonance. Under such conditions, we expect the gas to respond approximately hydrostatically in the $y$ direction,
in order to prevent the development of rapid horizontal velocities $u$ and $v$ at large $k_y$ (see equation (\ref{eq:v})).
That is,
\begin{equation}
p( x, z)  = - \rho_{0}(z) \Phi
\label{ph}
\end{equation}
and
 $p$ is a real function (in phase with real quantity $\Phi$). We  show that $y-$hydrostatic equation (\ref{ph}) is well satisfied  for $k_y h \gg 1$ in
Appendix A.

%Since $w$ has singular behavior near the resonance and $p$ does not, the advective term involving $w$ dominates
%the entropy equation there.
The torque density integral in equation (\ref{Ti}) can then be readily evaluated. 
We use equation (\ref{ph}) to determine that
\begin{equation}
\frac{d T}{dz} =  \pm \frac{ \pi^2}{4}\frac{\gamma-1}{\gamma} r^2 \rho_0(z) \Phi^2 \frac{N}{ A  \, c^2} 
%\label{T1}
\end{equation}
or
\begin{equation}
\frac{d T}{dz} =   \pm \frac{ \pi^2}{4} r^2 \rho_0(z) \Phi^2 \frac{N^3}{ A  \, g^2}.
\label{T1}
\end{equation}
Notice that the torque density is independent of the resonance width $\l_{\rm d}$ and azimuthal wavenumber $k_y$, provided that $k_y$ is large.

 The torque density derivation assumed that the gas behaves hydrostatically in the azimuthal direction, equation (\ref{ph}).
 As discussed in Appendix B,  this assumption breaks down in a region whose thickness is of order $1/k_y$ about the
 disk midplane, where the effects of the disk midplane boundary condition, equation (\ref{zbc}),  are important.
 This boundary condition requires the torque density $dT/dz$ to vanish at the disk midplane.
 The resulting torque is given by equation (\ref{dTdz2}) of Appendix B.

\subsection{Numerical Solution of Separable Equations \label{ns}}
\begin{figure}
\epsscale{1.0} \plotone{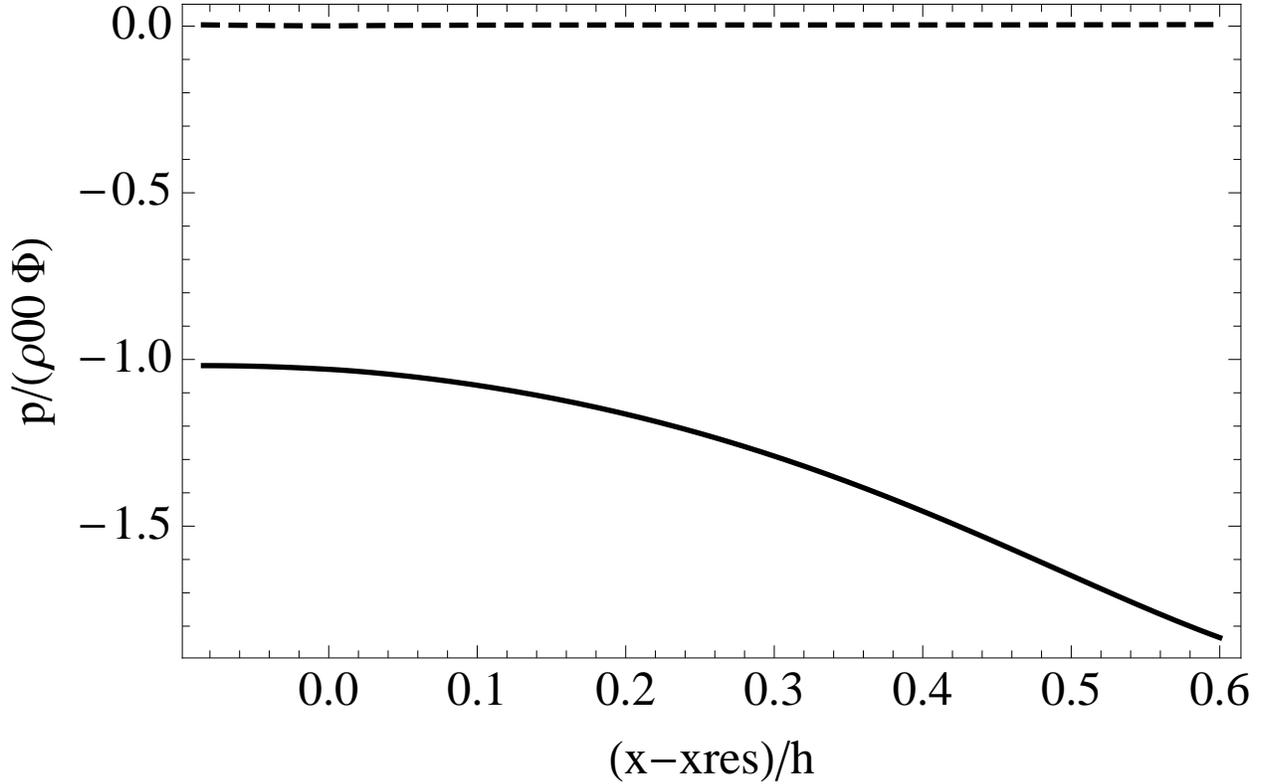} \caption{
Numerical solution of equation (\ref{peq1})  for the dimensionless pressure perturbation $P(X)= p(X)/(\rho_{00} \Phi)$ as a function of position $X=(x-x_{\rm res})/h$
relative to an outer buoyancy resonance. We take
$K_y=k_y h =5$ and $\gamma=5/3$. We apply the
$x$ boundary conditions of Section \ref{bc}
with inner boundary located at $x=0$ and the outer boundary  at $x=0.6 h+x_{\rm res}$.
The solid (dashed) line is the real (imaginary) part of $P$. Notice that $p \simeq -\rho_{00} \Phi$
near the resonance ($X=0$),
in agreement  with the hydrostatic approximation of equation (\ref{ph}) and 
the series solution (\ref{c0}) for large $K_y$. 
The resonance was
resolved with $\epsilon = 1 \times 10^{-8}.$
} \label{fig:px}
\end{figure}

In Appendix A, we obtain a separable equation for the dimensionless 
pressure perturbation $P(X,Z)= p(X) \exp{(-|Z|)}/(\rho_{00} \Phi)$ off the midplane
($|z| > 1/k_y$, away from the influence of the $z=0$ boundary),
with $X= (x-x_{\rm res})/h$ (a different $X$ than defined in Section \ref{TDcg}) and $Z=z/h$.
We solve  for pressure $P(X)$ given by equation (\ref{peq1}) numerically as a two point boundary value problem with
the $x$ boundary conditions that are described in Section \ref{bc}.
To determine this solution, we need to specify how to treat the singular $1/X$
term in the numerical integration across the resonance.
Following the procedure described in Section \ref{beq}, 
we resolve the singularity by replacing $1/X$ by $1/(X + i\, \epsilon)$ where $ 0 < \epsilon \ll 1$ is a dimensionless length scale 
defined as $\epsilon = \l_{\rm d}/h$
that we choose to be $\epsilon=1\times 10^{-8}$. We 
adopt values for  the adiabatic index $\gamma = 5/3$ and  $K_y= k_y h = 5$.
Fig.~\ref{fig:px} plots the results for $P(X)$.
Notice that the pressure perturbation $P(X)$ varies smoothly near the resonance,
and its value there agrees well with the value of -1 predicted by the
$y$-hydrostatic approximation (\ref{ph}), and as expected by
the series solution (\ref{c0}) for large $K_y$.

\begin{figure}
\epsscale{1.0} \plotone{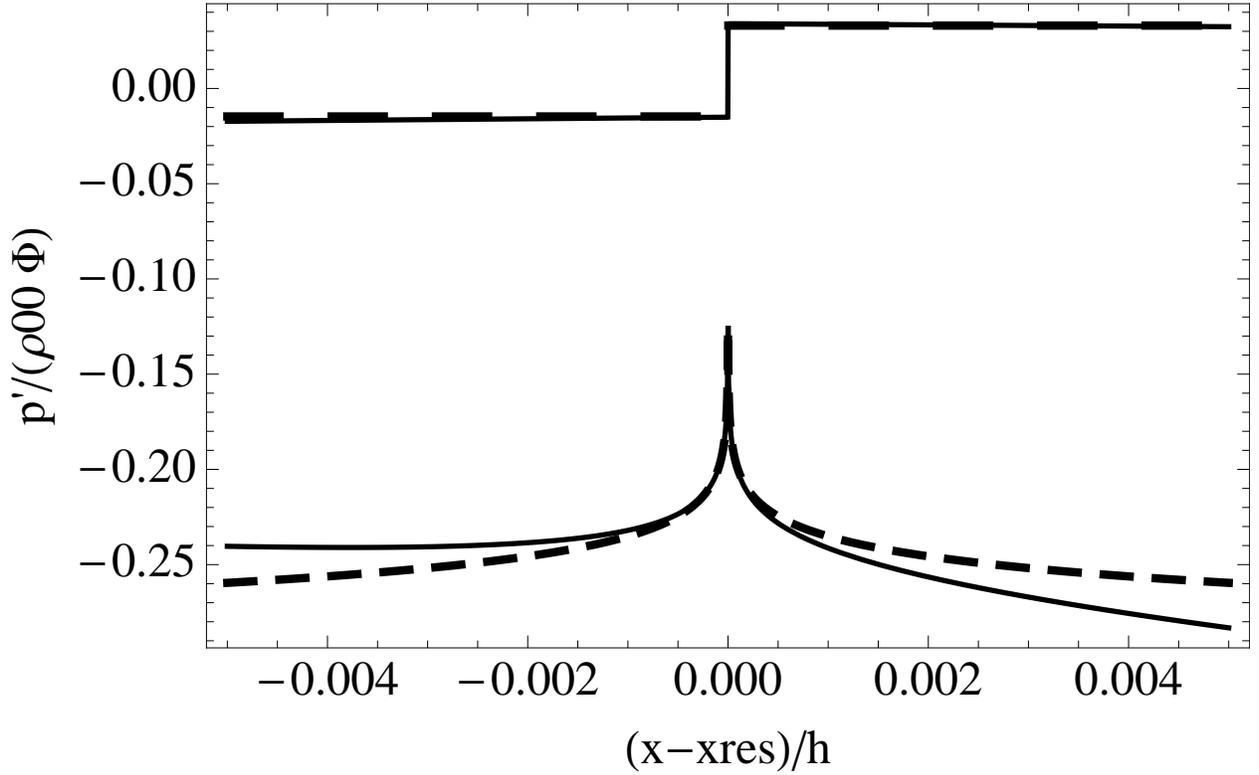} \caption{
Numerical solution for the dimensionless first derivative of the pressure perturbation $P(X)$ as a function of $X$ for the case described in 
Figure \ref{fig:px}. The upper (lower) solid line plots the imaginary (real) of part of $P'(X)$.
The upper (lower) dashed line plots the imaginary (real) part of the analytic solution near the resonance
given by equation (\ref{psing2}), together with equations (\ref{c3}) and (\ref{c1}), and  $d_1 = - 0.325+0.033i$. 
}  \label{fig:dpx}
\end{figure}

In Appendix A we determine the analytic properties of the buoyancy resonance.
We verify them with the numerical solutions for the separable equations.
Fig.~\ref{fig:dpx} shows the behavior of the numerical solution for $P'(X)$ near the buoyancy
resonance at $X=0$. %From equation (\ref{psing2}), we expect $P'(X) \sim c_1 \log{|X|} + i c_3 H(-X)$.
In agreement with the analytic calculation, equation (\ref{psing2}) of Appendix A, we  see that there is a jump in the imaginary part of $P'$ and a logarithmic singularity 
in the real part of $P'$. The dashed lines in Fig. ~\ref{fig:dpx} plot the analytic approximation to
$P'(X)$ near the resonance and show good agreement. 
%The constant term $d_1$  in equation (\ref{psing2}) is $d_1 = - 0.325+0.033i$ in this case.

\begin{figure}
\epsscale{1.0} \plotone{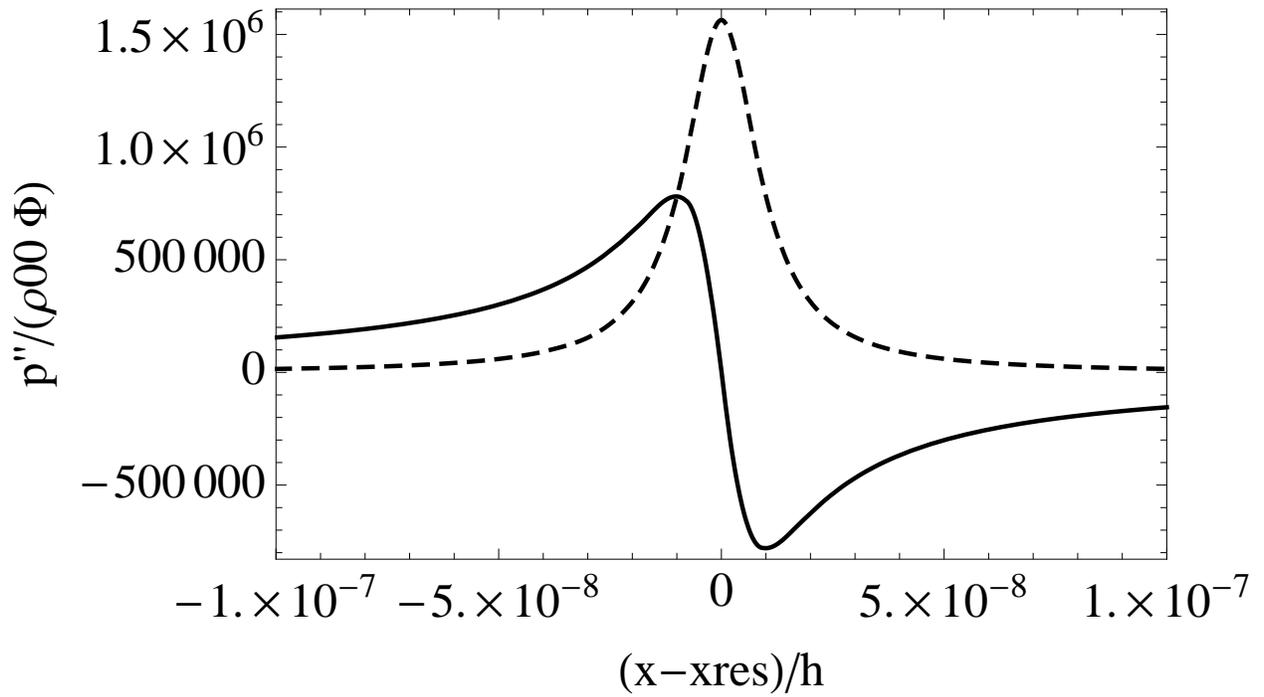} \caption{
Numerical solution for dimensionless second derivative of the pressure perturbation $P(X)$ as a function of $X$ for the case described in 
Figure \ref{fig:px}. The solid (dashed) line is the real (imaginary) part of $P''(X)$. The resonance was resolved with a damping
parameter $\epsilon = 1 \times 10^{-8}$ that determines the $X$ scale of the curves, as discussed in Section \ref{ns}.} \label{fig:d2px}
\end{figure}

Fig.~\ref{fig:d2px} shows the behavior of the numerical solution for $P''(X)$ near the buoyancy
resonance at $X=0$. From equation (\ref{psing1}) of Appendix A, we expect $Im(P''(X))$ to be of the form of $\delta(X)$,
for Dirac delta function $\delta$, as is 
consistent with the plotted function. As discussed in Appendix A, this out of phase delta function contributes
to the out of phase density response that in turn contributes to the localized torque at the resonance $X=0$.

The leading variations in $X$ for $ |X| \ll 1$ and weak damping ($0< \epsilon \ll 1$) 
 of the various physical quantities at the buoyancy resonance for $K_y \gg 1$ and
$|Z| \gg 1/K_y$ are
given by 
\begin{eqnarray}
Re(u) &\sim& \arctan{(X/\epsilon)} \sim H(X), \label{simi}\\
Im(u) &\sim&  \log{(X^2 + \epsilon^2)} \sim \log{(|X|)}, \\
Re(v) &\sim&  \log{(X^2 + \epsilon^2)} \sim \log{(|X|)}, \\
Im(v) &\sim&  \arctan{(X/\epsilon)}  \sim H(X), \\
Re(w) &\sim& \frac{\epsilon}{X^2+\epsilon^2} \sim \delta(X), \\
Im(w) &\sim& \frac{X}{X^2+\epsilon^2} \sim 1/X, \label{Imw}\\
Re(\rho) &\sim& \frac{X}{X^2+\epsilon^2} \sim 1/X, \label{Rerho}\\
Im(\rho) &\sim&  \frac{\epsilon}{X^2+\epsilon^2} \sim \delta(X),\\
Re(p) &\simeq& -\rho_0 \Phi =  -\rho_0 \Phi, \\
Im(p) &\sim& X +  b_1\, X \arctan{(X/\epsilon)}  \sim X +  b_2\, X H(X), \label{simf}
\end{eqnarray}
where $H(X)$ is the Heaviside step function, $b_i$ are real constants, and coefficients have been omitted in the $\sim$ relations. 
The third column in equations (\ref{simi}) -   (\ref{simf}) contains the functional forms in the limit of $\epsilon \rightarrow 0$
at fixed $X$.

Unlike the case of Lindlbad resonances,
there are no radially or vertically propagating waves launched. 
Unlike the case of the previously studied disk resonances, the resonance width in $x$ is $\l_{\rm d}$ that
is unrelated to the gas sound speed. The width is due to the effects of damping forces and radiative diffusion.

\subsection{Comparison of Analytic Model With Simulations \label{sim}}

\begin{figure}
\epsscale{1.0} 
\plotone{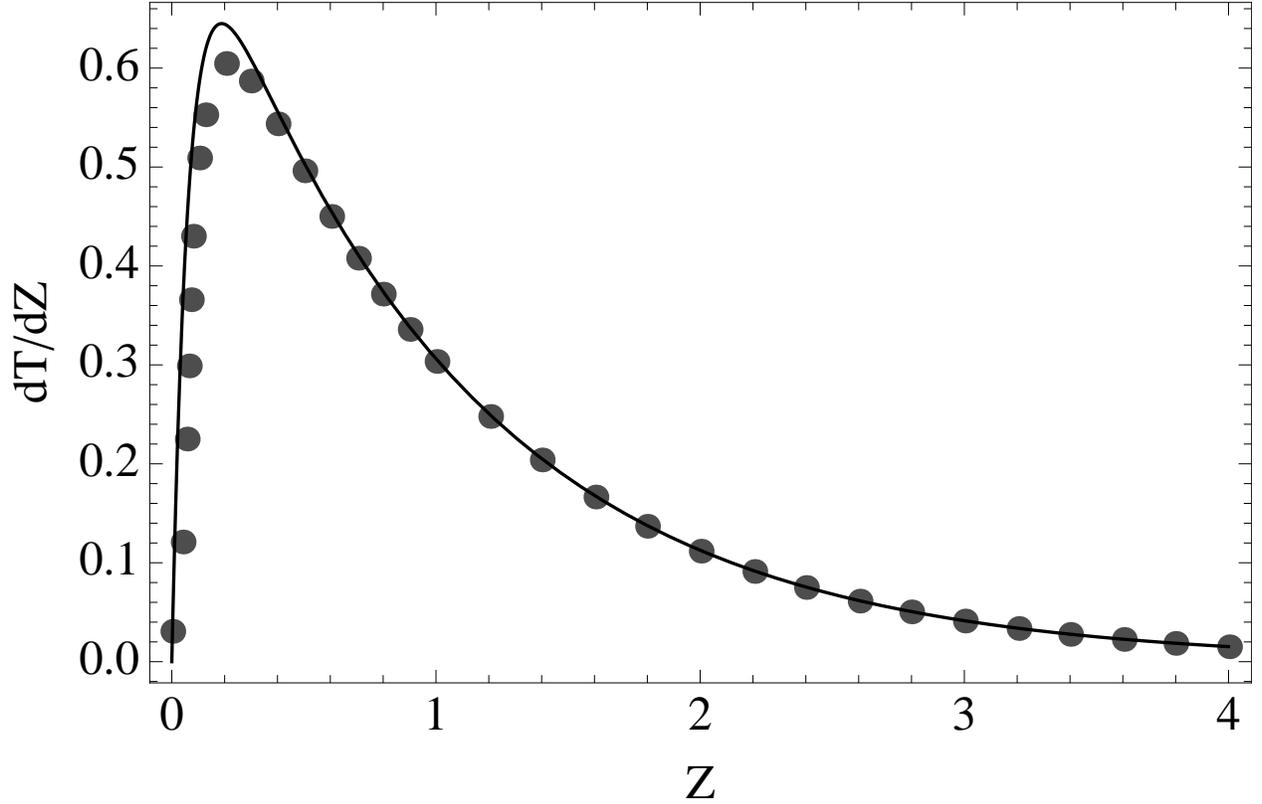} \caption{
Dimensionless torque density distribution $dT/d Z$, normalized by $r^2 h \rho_{00}  \Phi^2 /c^2$, 
at an outer buoyancy resonance as a function of height above the disk midplane $Z=z/h$  
for Case (a) 
described in  Section \ref{TDcg} that has
 constant vertical gravity and simplified potential.
The line plots the analytic expression (\ref{dTdz2}) 
and the dots are the results of 3D nonlinear
shearing
box simulations with the Athena code that are evaluated through equation (\ref{dTdZ}). } \label{fig:dTdz2}
\end{figure}

Based on Athena simulation results, we determined the  torque density $dT/dZ$ for Case (a)  that has constant $\Phi$, as described
in Section \ref {TDcg}. Along the lines
of  equation (\ref{dTdXdZ}), we calculated the dimensionless torque density $dT_{\rm s}/dZ$  for an outer resonance
from the results of simulations
 by integrating the torque density in both the $x$ and $y$ directions
at various heights $Z$ above the disk midplane
  \begin{equation}
  \frac{d T_{\rm s}}{dZ} =   \frac{\pi K_y} {\Phi_{\rm s}} \, \frac{1}{2 Y_{\rm b}} \int_{-Y_{\rm b}}^{Y_{\rm b}} \int_{0}^{X_{\rm b}}  \rho_{\rm s} (X,Y,Z) \,  \, \sin{(K_y Y)} \, dX  \, dY,
  \label{dTdZ}
  \end{equation}
  where again $\rho_{\rm s}$ is the dimensionless gas density in units of the 
  unperturbed midplane disk density $\rho_{00}$,   $\Phi_{\rm s} = \Phi/c^2$ is a dimensionless real constant, and $X_{\rm b}$ and $\pm Y_{\rm b}$
  are the dimensionless locations of the outer $x$ boundary and both $y$ boundaries of the simulations,
  respectively.
 The dimensional torque density $dT/dZ$ is obtained by multiplying $dT_{\rm s}/dZ$
  by $r^2 h \rho_{00}  \Phi^2/c^2$. 
As seen in the Panel (a) of Fig.~\ref{fig:tdens}, the torque density was confined to a region of small radial extent
centered on the buoyancy resonance.
Fig.~\ref{fig:dTdz2} compares the nonlinear simulation results for $dT_{\rm s}/dZ$ with predictions based on the analytic expression
(\ref{dTdz2}) and shows good agreement. 
%The largest
%discrepancy occurs near the peak that we have found to have some dependence
%on the form of artificial viscosity

\section{MODEL WITH VARIABLE VERTICAL GRAVITY AND SIMPLE AZIMUTHAL FORCING  \label{safvg} }

\subsection{Torque}

We extend the constant gravity and simple potential model of Section \ref{saf} to the more realistic 
case
in which the vertical gravity varies linearly in $z$, as expected in a thin disk. We apply the simple potential 
given by equation (\ref{eq:Phic}).
Vertical hydrostatic balance implies that the unperturbed disk satisfies
\begin{eqnarray}
p_0(z) &=& p_{00} \exp{(-z^2/(2 h^2))}, \label{pv0}\\
\rho_0(z) &=& \rho_{00} \exp{(-z^2/(2 h^2))}, \label{rhov0}
\end{eqnarray}
where 
\begin{equation}
h = c/\Omega
\end{equation}
with constants $c$, $h$, $p_{00},$ and $\rho_{00} = p_{00}/c^2.$ 
We also have
\begin{equation}
g(z) = \Omega^2 \, z
\end{equation}
and
\begin{equation}
N(z) =\sqrt{ \frac{\gamma-1}{\gamma}} \Omega \, \frac{|z|}{h}.
\end{equation}

Based on the torque density expression (\ref{T1}), we postulate that the torque in the variable
gravity case is given by
\begin{equation}
\frac{d T}{dz} =   \pm \frac{ \pi^2}{4} r^2 \rho_0(z) \Phi^2 \frac{N^3(z)}{ A  \, g^2(z)}.
\label{Tv1}
\end{equation}
This torque density expression follows from the application hydrostatic condition (\ref{ph})  to
the torque equation (\ref{Ti}) with variable gravity. For small $z$, the torque density varies linearly with $z$.

%The torque on the gas both above and below the disk midplane is given by
%\begin{equation}
%T =   \pm \frac{ \sqrt{2} \pi^{3/2}}{8} \left(\frac{\gamma-1}{\gamma} \right)^{3/2} \frac{r^2 \Sigma_0 \Phi^2}{ A \Omega }.
%\label{Tv1}
%\end{equation}

The integrated buoyancy torque for each $k_y > 1/h$ is then given by
\begin{equation}
T =  \pm \frac{\sqrt{2} \pi ^{3/2}}{4 } \left(\frac{\gamma -1}{\gamma } \right)^{3/2}   \left( \frac{r}{h} \right)^2 \Sigma  \frac{\Phi
   ^2}{ A \Omega},
\end{equation}
independent of $k_y$.
 The Lindblad torque has a much different
dependence on wavenumber $k_y$. 
We compare the buoyancy torque $T$ to the Lindblad torque $T_{\rm L}$ subject to the same potential
given by equation (\ref{eq:Phic}). We consider wavenumbers near the torque cutoff, $k_y r \sim r/h \gg 1$
and have that
\begin{equation}
T_{\rm L} =  \mp \frac{4 \pi ^2 \, k_y^2 r^2 \Sigma\,  \Phi ^2}{3 \Omega ^2},
\end{equation}
which follows from equation (13) of \cite{GT80} for a Keplerian disk.

The ratio of the torques in a Keplerian disk with $\gamma=5/3$ is given by
\begin{equation}
 \frac{T_{\rm L}}{T} =  5 \sqrt{5 \pi} (k_y h)^2 \simeq 20  (k_y h)^2.
 \label{TLT}
\end{equation}
Since we are considering wavenumbers with $k_y h \sim 1$,
we see that the Lindblad torque is much stronger.

Zhu et al (2012) reported that the one-sided (inside or outside corotation) total buoyancy torque in the case of the planet potential was tens of percent of the one-sided
total Lindblad
torque and was then  much stronger in a relative sense than suggest by equation (\ref{TLT}).
This difference is likely due to the use here of a potential that is independent of $x$ and $z$.
The buoyancy torque in the point mass case may be getting stronger potential contributions that lie closer to
the planet than the Lindblad torque.  
In addition, the Lindblad torque declines above its torque cutoff, for $k_y \ga 1/h$.
In the point mass case, the Lindblad and buoyancy torques likely vary differently with $k_y$ for wavenumbers above the Lindblad
torque cutoff. Such differences may explain why the buoyancy torque (integrated over $k_y$) is relatively stronger in the point mass case than the estimate here suggests.

\subsection{Linear Numerical Calculation \label{simv}}

In Section  \ref{sim} we showed that the torque obtained by linear theory
agrees well with that determined by nonlinear simulations.
Unlike the constant vertical gravity case described in Section \ref{ns}, we cannot obtain separable solutions
to the linearized equations.
Instead, we determine the torque in the variable gravity case through a
numerical solution to the  linearized equations by means of a Fourier method in $x$.
This method is a 3D extension of the sheared coordinate approach taken by \cite{GT78}.
We describe the method in Appendix C.

Equations (\ref{eq:u2v}) - (\ref{eq:s2v}) describe the dynamics 
in terms of the $x$ Fourier transforms of physical quantities $q(x, z)$ to quantities $\hat{q}(\tau, z)$.
The equations are expressed in terms of a time-like coordinate $\tau$ and vertical coordinate $z$.
These equations 
were nondimensionalized by setting $\Omega=1$, $h=1$, $\rho_{00} =1$ and  following the dimensionless variable notation
of Section \ref{TDcg}. 
We integrated these equations by means of the Method of Lines in Mathematica
with the implicit Runge-Kutta option for various values of $K_y$ using
 $\nu_w = 0.001$,   $\hat{\Phi}=0$ (see explanation
above equation (\ref{eq:ubc})). The integration extended from $\tau= \tau_{\rm s}=0$ to $\tau=\tau_{\rm f} = 140/\sqrt{K_y}$
and $Z$ from 0 to 4.
Boundary conditions (\ref{eq:ubc}) - (\ref{eq:pbc}) were applied with $\Phi=1$.
The results were transformed back to $q(X, Z)$ and the torque density, normalized by $r^2 h \rho_{00} \Phi^2/c^2$, was determined
from the integral 
\begin{equation}
\frac{d T}{d Z} = -\pi K_y \int_{0}^{X_{\rm o}} Im(\rho (X,Z)) \,dX
\end{equation}
in dimensionless units.
The width of the peak in the density profile in $X$ in the numerical calculation
is determined mainly by the length of the $\tau$ integration, $\tau_{\rm f}$. 
The upper limit of the torque integral $X_{\rm o}$ was  typically chosen to be $2 X_{\rm res}$
that is far enough away from the resonance that the resonant density perturbation
is small.

\begin{figure}
\epsscale{1.0} 
\plottwo{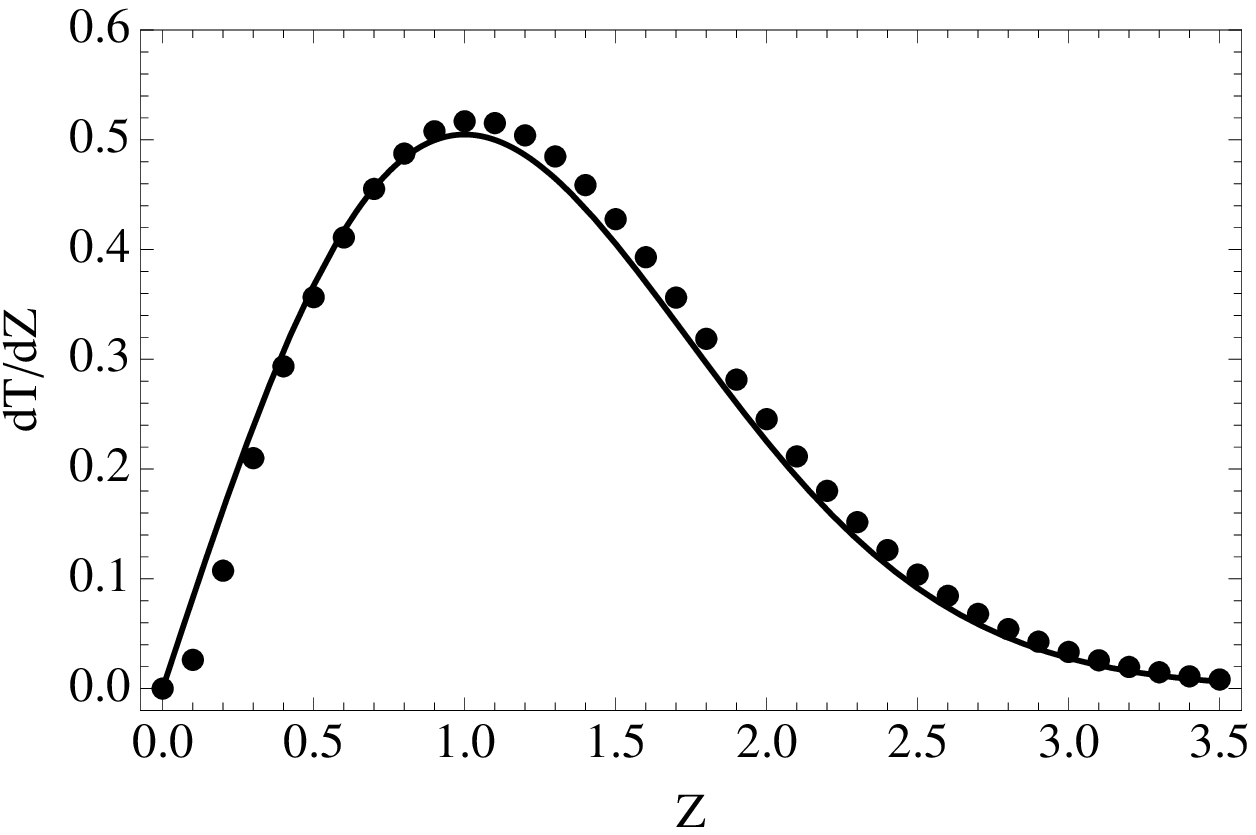}{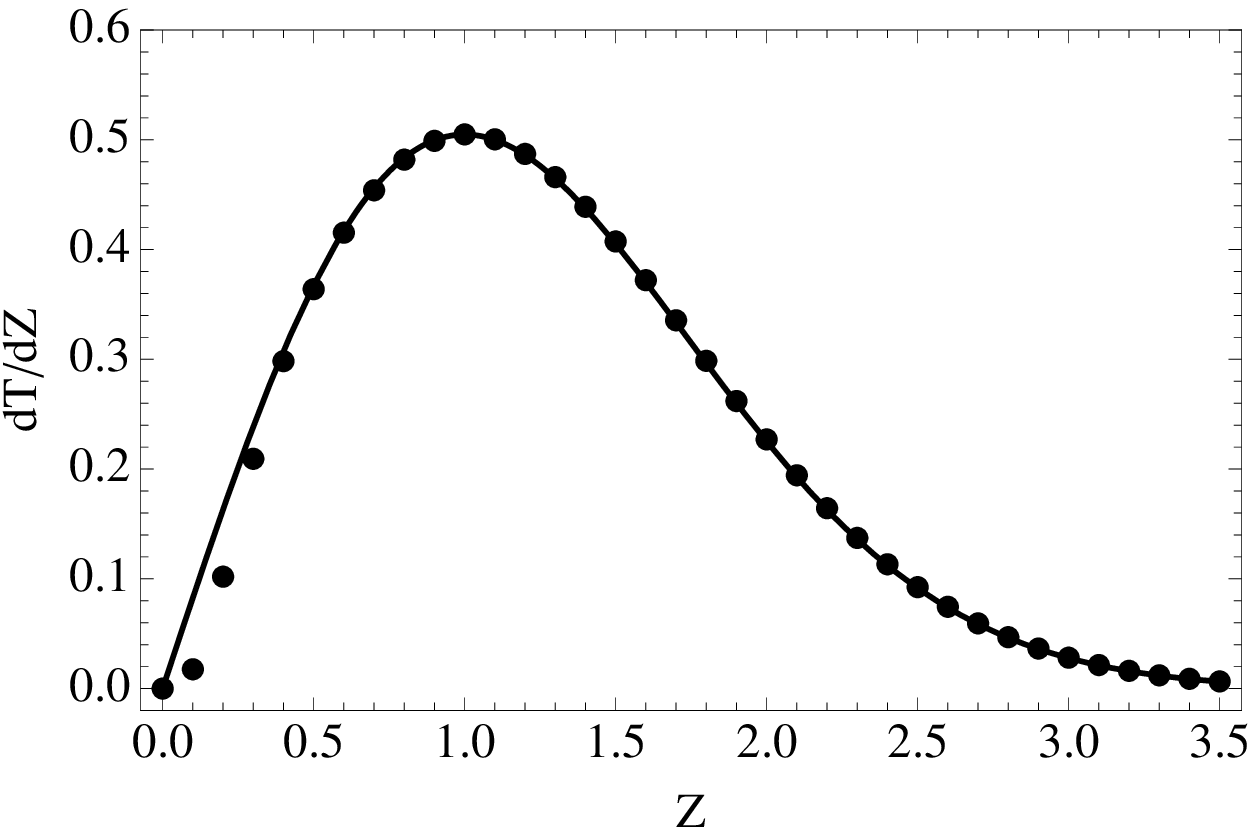} \caption{
Torque density distribution $dT/d Z$ normalized by
$r^2 h \rho_{00}  \Phi^2 /c^2$
at an outer buoyancy resonance 
as a function of height above the disk midplane $Z = z/h$
for  models with 
$k_y h =5$ (left panel) and $k_y h =10$ (right panel).
The vertical gravity varies linearly with $z$ and the perturbing potential
is given by equation (\ref{eq:Phic}). 
The line plots analytic expression (\ref{Tv1}) and the dots are the results of numerical linear 3D shearing
box calculations that are based on a Fourier method in $x$ described in Appendix C.  } \label{fig:dTdzv}
\end{figure}

In Fig.~\ref{fig:dTdzv} we compare the results of this calculation with the
torque density equation (\ref{Tv1}) for  cases with $K_y=5$ and 10 and $\gamma=5/3$.
As expected, the torque distribution is independent of $K_y$ for this potential.
The agreement is very good.

\section{DISCUSSION \label{disc}}

Buoyancy resonances have been previously analyzed in the context of high mass stars
that are tidally perturbed by a companion \citep{Z75, GN89}. A buoyancy  resonance occurs near the 
outer parts of the stellar convective core where the buoyancy
frequency changes from low values in the core to high values in the
radiative envelope. Within this region, the buoyancy frequency matches that of 
the tidal forcing. The response of the star is to launch buoyancy waves (g mode waves)
that propagate towards the stellar surface where the waves damp and act to bring about spin synchronization
with the binary orbit. 

As we see in this paper, the adiabatic response of a disk to tidal forcing is evidently quite different.
In the axisymmetric case of an isothermal disk,  buoyancy waves (g modes) can be excited at a Lindblad resonance
\citep{Bate02}. They propagate on the same side of the Lindblad resonance as the fundamental
mode propagates which is not the region where the buoyancy resonance discussed here
is found.  For a vertically truncated isothermal disk, they accumulate near the disk upper boundary. We do not
find evidence for these waves in the numerical results presented here at small $|x| < h$.

Low frequency axisymmetric waves in the form of r  (rotation-dominated) modes exist and are affected by buoyancy that
confines them near the disk midplane (see Fig. 11 of \cite{LP93}).
These modes can also be excited at a Lindblad resonance and propagate toward corotation ($x=0$),
although they are only weakly excited there \citep{Bate02}.

To understand the star-disk difference,
consider  low frequency modes of given $k_y$  in the region where 
wave frequency $|\omega|$ is smaller than epicyclic frequency $\kappa$,
which is the region where the buoyancy resonances described in this paper reside.
In a region such that $k_y \ll |k_x|$,
the WKB dispersion relation for the disk pressure perturbation is 
\begin{equation}
k_x^2    =  \frac{\kappa^2 - \omega^2}{\omega^2 - N^2(z)} \,  k_z^2, 
\label{dr}
\end{equation}
where $\omega$ is the wave frequency
and 
\begin{equation}
\omega= 2 A k_y x
\end{equation}
\citep{VD90, G93}.
This relation can be easily derived from equation (\ref{peq}).
%In determining modes, $k_x$ is taken to be independent of $z$, while $k_z$ depends on $x$ and $z$.
We see then from
equation (\ref{dr}) that vertical propagation is possible ($k_z^2 >0$)
only for low values of the buoyancy frequency, $N^2< \omega^2$. For a disk
with $N \propto z$,  such a wave can propagate vertically
only in the region near the disk midplane as an r mode.

Consider the stellar case.
To see how waves can be launched towards  a stellar surface,
we adapt equation (\ref{dr}) to the case of a star by noticing that $k_z$ in a disk
describes phase variations along the direction of the buoyancy gradient. In the case of a star,
the buoyancy gradient is in the (spherical) radial direction, and so we identify $k_z$ with $k_r$.
We identify  the square of the wavenumber perpendicular to the buoyancy gradient
with $k_x^2  \sim \ell (\ell+1)/r^2$ for the spherical harmonic of  order $\ell$ that is associated
with the tidal field. We disregard the stellar rotation and set $\kappa=0$.
The tidal forcing frequency due to the companion star is $\omega$.
We then have that
\begin{equation}
k_r^2 = \frac{\ell (\ell+1)}{r^2} \frac{N^2- \omega^2 }{\omega^2}. 
\label{drs}
\end{equation}
Radial wave propagation is possible (i.e., $k_r^2 >0$) in the radiative outer
layers of a star outside the buoyancy resonance  where $N^2(r) > \omega^2$ (cf. equation (15) of \cite{GN89} with $\ell=2$).
Waves can be launched at the resonance  where $N^2(r) \simeq \omega^2$ because the long wavelength of the wave $\sim 1/k_r$ can match the
spatial scale of the slowly varying tidal field. The effects of rotation modify the dispersion relation
by changing $\omega^2$ in the denominator on the right-hand side of equation (\ref{drs})
to $\omega^2-\kappa^2$.
Therefore, one major difference between the disk case investigated here and stellar case
is due to the effects of rapid rotation, $\kappa$, in the low frequency disk case.  

Consider possibility that r modes are launched
at buoyancy resonances in disks. Waves are launched at resonances if
there is a strong overlap between the spatial form of the wave with that of the forcing.
Typically that occurs at wave turning points where the long wavelength form of the wave
matches the form of the relatively slowly varying potential. For example, from equation (\ref{dr}),
 $k_x$ is small near a Lindblad resonance where $\omega = \pm \kappa$.
This locally long wavelength permits a strong coupling with the tidal potential that
results in the excitation of a wave. 
Although vertical wave turning points for r modes lie on the buoyancy resonance plane  
($k_z=0$ at $\omega=N$ in equation (\ref{dr})),  the radial wavenumbers $k_x$ are large,
as follows from equation (58) of  \cite{LP93} with dimensionless frequency $F= 2 A x k_y/\Omega$ for $|F| \ll 1$.
(Note: \cite{LP93} used the term g modes for what are r modes, see also
 \cite{O98}.) The r mode becomes more confined vertically as it approaches corotation. Its 
 vertical wavenumber $k_z$ at the midplane and radial wavenumber 
 $k_x$ both grow to very large values near corotation ($x=0$). 
Therefore,
the r mode structure  
may not generally  match the spatial form of the potential along the buoyancy resonance.
This lack of matching may explain why r modes are  not excited at buoyancy resonances.

Differential equation (\ref{peq1}), which we obtained for the  pressure perturbation near a buoyancy resonance in the constant gravity case,
is consistent with dispersion relation (\ref{dr}). In that case, $N$ is constant,
$k_z^2 = -1/h^2$ (see equation (\ref{psf})), and near the buoyancy resonance 
$N^2 - \omega^2 \propto x-x_{\rm res} \propto X$. We then have
that $k_x^2 \propto 1/X$ which covers the leading order terms on the left-hand side
of equation (\ref{peq1}), identifying $P''(X)$ with $-k_x^2 P(x)$. However, the WKB approximation
does not provide much insight into the properties of a buoyancy resonance because it does not involve waves.

\cite{LP93} determined the structure of the disk modes  by using what could be called the "waveguide" model.
In that model, the amplitude variations in the $x$ direction are assumed to occur slowly compared to phase variations. 
For perturbations induced by the buoyancy resonance, that assumption does not apply. For the case of vertically
varying gravity, amplitude variations along both the $x$ and $z$ directions are rapid and comparable.

%We have found
%some evidence for surface gravity wave excitation in the numerical solutions discussed in Section \ref{safvg}.
%Their strength seems to depends sensitively on the height of the disk upper boundary $z_{\rm u}$
%and appears weak for the $z_{\rm u}$ large.
%We do not pursue this issue further here.

\section{SUMMARY \label{sum}}

The interaction between a planet  and  a disk that responds adiabatically
can be understood qualitatively in terms of a 3D impulse delivered to the gas
as it passes by the vicinity of the planet \citep{Zhu12}. The impulse generates a wake
whose density is affected by the buoyancy of the gas. The wake in turn causes a planet-disk torque.

To understand the physical nature of the response of the gas, we have examined its 
nonaxisymmetric response to a potential that has a single azimuthal wavenumber $k_y$. 
We analyzed the role of buoyancy resonances in a disk whose
unperturbed vertical structure is isothermal  and is subject to adiabatic perturbations. 
A vertically displaced fluid element undergoes vertical free oscillations
at the buoyancy frequency $N$. 
Although the gravitational forcing we have adopted was simplified to be purely azimuthal,
this forcing induces nonaxisymmetric vertical pressure and buoyancy forces with azimuthal wavenumber $k_y$ 
that are stationary in the frame of the potential.
Fluid elements move azimuthally through this vertical force field at velocity $2 A x$ due
to the disk shear. 
The vertical forcing frequency on the fluid elements is then $2 A x k_y.$
A buoyancy resonance occurs where
the absolute values of the  free ($N$) and forcing frequencies ($2 A x k_y$) match, as given by
equation (\ref{xres}).  
The resonance
leads to a nonaxisymmetric density response that contains a contribution that is out of phase
 with respect to the potential and results in a torque.

To carry out the analysis,
we first considered the case of a disk with constant vertical gravity that is subject to a
simple perturbing potential  given by equation (\ref{eq:Phic}).
The resonance in this case lies along a plane that is perpendicular to the disk plane, at some $x_{\rm res}$
(independent of $z$) given by equation  (\ref{xres}).
We obtained an analytic linear
description of the structure of the resonance.
The effects on the gas are highly localized to the plane of the resonance. The results show that the 
localized induced motions
cause localized density perturbations and a torque (see Fig.~\ref{fig:dTdz2}).
This result confirms the existence of the buoyancy torque found in simulations by Zhu et al (2012).
 
 The pressure perturbation is everywhere smooth, but singular behavior occurs
in its second radial derivative that leads to a contribution to the torque (Figs. \ref{fig:px} - \ref{fig:d2px}) from the radial derivative of the radial velocity. 
The width of the torque region is controlled by damping processes and not gas pressure.
The resonance does not result in radially or vertically propagating waves.  
 
 We then considered the case of variable vertical gravity with the same simple perturbing potential.
 Unlike the case of disk resonances previously studied, the buoyancy resonances in this case lie
on tilted planes $x_{\rm res} \propto z$. 
 We obtained an analytic formula for the torque density, 
equation (\ref{T1}) that agrees well with numerical calculations (see Fig. \ref{fig:dTdzv}).
%The torque contribution from the Lindblad resonance in this region is very small.
The buoyancy resonance exerts a torque over a region 
that lies radially closer to the corotation radius  than the Lindblad resonance.

 Bouyancy resonances have then very different properties from  
 Lindblad resonances. They do not result in vertically or horizontally propagating waves.
 Their width is not determined by the gas sound speed, but instead by damping.
  
The current analysis has several limitations that could be overcome in future studies.
The potential was taken to be of a very simple form in order to investigate
the existence and basic analytic properties of a buoyancy resonance. But a more realistic
potential would provide a more accurate description in a linearized model.
The analysis presented here assumed that the gas behaves
adiabatically. The radiative transfer of heat between the resonant region and its surroundings
can reduce the buoyancy force
on perturbed gas and so weaken the resonance for sufficiently low values of the disk optical depth. 
In addition, this heat exchange plays a role in determining the resonance width.

Since the torque is confined to a thin layer, the resonance may saturate (weaken)
by feedback effects that may act to change the local disk structure.  
Such effects cannot be studied by linear theory.
The analysis thus far has been limited to shearing boxes.  The shearing box does not
describe the corotation resonance and its possible interaction with the buoyancy
resonance.
The effects of the buoyancy torque on the net migration rate and in particular its direction (inwards or outwards)
depend on a competition between the inner and outer buoyancy torques that
in turn depends on gradients of disk parameters. The determination of the outcome
requires going beyond the shearing box approximation.

%\section*{ACKNOWLEDGMENTS}

\acknowledgments

We benefitted from useful discussions with Gordon Ogilvie, Roman Rafikov, and Jim Stone. 
SHL acknowledges support from NASA Origins grant NNX11AK61G.
 ZZ acknowledges support by NASA through Hubble Fellowship grant HST-HF-51333.01-A 
 awarded by the Space
Telescope Science Institute, which is operated by the Association of Universities for Research in Astronomy, Inc.,
for NASA, under contract NAS 5-26555. 
All simulations were carried out using computers supported by the Princeton Institute of Computational Science and Engineering 
and Kraken at National Institute
for Computational Sciences through XSEDE grant TG-AST130002.

 \appendix
\section{PRESSURE PERTURBATION FOR CONSTANT VERTICAL GRAVITY AND SIMPLE AZIMUTHAL FORCING \label{AppA}}

We examine the behavior of pressure near the buoyancy resonance and
verify the $y$-hydrostatic approximation of equation (\ref{ph}) for
the perturbing potential of the form given by equation (\ref{eq:Phic}) in the limit of $k_y h \gg 1$.
Above the disk midplane, for a given $k_y$, 
the gas dynamical equations (\ref{eq:u}) - (\ref{eq:s}), 
together with equations (\ref{p0}) and (\ref{rho0}), can be combined
to provide a single equation for the pressure perturbation $p$ that is given by
 \begin{eqnarray}
 \nonumber 
& & \frac{\kappa^{2}-(2 A k_y x)^{2}}{(2 A k_y x)^{2}-N^{2}}  \left ( \partial_z^{2} p  +
\frac{\partial_z p}{h}   +
\frac{ p (\gamma-1)}{\gamma^2 h^2} 
 %  -
% \rho_0 \partial_z f_z+ \frac{\rho_0 f_z }{\gamma h}
 \right ) \\ \nonumber 
&=&
 \partial_x^{2} p  +
 \partial_x p \,  \frac{8 A^2 k_y^2 x }{\kappa^{2}-(2 A k_y x)^{2} }   
 + p  \left(\frac{8 A \Omega k_y^2 }{\kappa^{2}-(2 A k_y x)^{2} } -  k_y^2     -
\frac{ ( \kappa^{2}-(2 A k_y x)^{2})}{\gamma c^2} \right) \\
  & &+  k_y^2 \,  \rho_{0}(z)\,  \Phi \left(\frac{8 A \Omega }{\kappa^{2}-(2 A k_y x)^{2} } -  1 \right).
  \label{peq}
 \end{eqnarray}
 
 For the case of constant vertical gravity above the disk midplane and the simple
 potential given by equation (\ref{eq:Phic}),
 equation (\ref{peq}) is separable in $x$ and $z$.
 The pressure perturbation is of the form
 \begin{equation}
p(x,z) = p(x) \exp{(-|z|/h)}.
\label{psf}
\end{equation}
 We apply a change to dimensionless variables 
 \begin{eqnarray}
 \nonumber
  X &=& \frac{x - x_{\rm res}} {h},\\
   \nonumber
  Z  &=& \frac{z}{h},\\ \nonumber
  P(X) &=& \frac{p(X)}{\rho_{00} \Phi}, \\
  K_y &=& k_y \, h
  \end{eqnarray}
  and obtain an ordinary differential equation of the form
   \begin{equation}
 \nonumber 
s_2(X) \, P''(X) +  s_1(X) \, P'(X) + s_0(X) \, \frac{P(X)}{X} = s_3(X),
  \label{peq1}
 \end{equation}
 where $s_i(X)$ are real polynomials in $X$ that are nonzero at the resonance, that is 
 $s_i(0) \ne 0.$ %Equation (\ref{peq1}) is independent of scale height $h$.

  Close to the buoyancy resonance (small $|X|$), 
  there is a $1/X$ singularity multiplying $P(X)$.
  We expect this term  to balance the term involving the highest order derivative in $X$, that is $P''(X)$.
  The reason is that this highest derivative term is most sensitive to the structure of $P(X)$.
If we assume that  $P''(X) \sim O(1/X)$, then $P \sim O(X \log{X})$ for $X>0$.
We then consider the following series for $P(X)$ for small $X>0$,
\begin{equation}
P(X) = c_0 + (c_1 X + c_2 X^2)\log{(X)} + d_1 X + d_2 X^2,
\label{psing}
\end{equation}
where $c_i$ and $d_i$ are coefficients to be determined.  We neglect higher order terms in $X$.
Notice that $p$ is smooth in $X$ and $\displaystyle \lim_{X \to 0+} P=c_0$. The first derivative of $P$
contains a logarithmic singularity and
its second derivative  contains a $1/X$ singularity at $X=0$. %(The expansion can be extended to negative
%$x$-values, but that extension provides no further information about the torque.) 

Substituting expansion (\ref{psing}) into equation (\ref{peq1}) for $P$,
we obtain in lowest order (order $1/X$) an equation of the form
\begin{equation}
c_0 = a_1 c_1,
\label{c0c1}
\end{equation}
where $a_i$ used here and below are constant terms that depend on   $A$, $\kappa$, $N$, and $\Omega$.
In next order, we obtain an equation of the form
\begin{equation}
c_0 + a_2 + a_3 c_1+ a_4 c_2 + a_5 d_1 + a_6 d_2+ (a_7 c_1 + a_8 c_2 ) \log{(X)} = 0,
\end{equation}
where $a_2$ arises from the inhomogeneous term.
We then require the coefficient of $\log{(X)}$ to vanish and  the sum of
the other terms to vanish. Namely,
 \begin{equation}
c_0 +a_2 + a_3 c_1+ a_4 c_2 + a_5 d_1 + a_6 d_2 =0
\label{c0c1phic2}
\end{equation}
and
\begin{equation}
a_7 c_1 + a_8 c_2  = 0.
\label{c1c2}
\end{equation}
We then solve the three equations (\ref{c0c1}), (\ref{c0c1phic2}), and (\ref{c1c2})
for three unknowns $c_0$, $c_1$, and $c_2$ in  terms of $d_1$ and $d_2$.
These two $d_i$ parameters are 
due to the two boundary conditions in $X$ for this second order equation in $X$.
%The torque is proportional to $c_0$ for the case of no vertical gravitational forcing (see equation (\ref{Ti})).

In the limit of large $K_y$ or equivalently small $X_{\rm res} = x_{\rm res}/h$,
we find
\begin{equation}
c_0 = \frac{p(x_{\rm res})}{\rho_{00} \Phi} = -1 + K_y^{-1}  q_1 d_1 +  K_y^{-2}  q_2 ( d_2 + q_3) + O(K_y^{-3})
\label{c0}
\end{equation}
where $q_i$  are
dimensionless order unity coefficients that depend on $A$, $\kappa$, $N$, and $\Omega$. 
In the large $K_y$ limit, we expect the terms on the right-hand side of equation (\ref{c0}) involving $K_y$ to vanish.
In this limit,
equation (\ref{c0}) then implies equation (\ref{ph}) that in turn implies the torque given by equation (\ref{T1}). 
%We generally expect that the boundary conditions typically do not require that $d_1$ increase linearly with $k_y$
%or $d_2$ to increase as $k_y^2$. To do so would imply that the perturbed pressure vary
%by order unity amounts over the small distance $|X_{\rm res}|$ between resonance and corotation that varies
%inversely with $k_y$.
%Smooth solutions then require that the terms involving $d_i$ can be ignored in the high $k_y$
%imit, compared to the $k_y$-independent term $-\rho_{00} \Phi$.
In practice, we find that the coefficient of the $K_y^{-1}$ is typically small.

Notice that we are assuming that the $d_i$ terms do not increase substantially with $K_y$.
To do so would imply that the  pressure perturbation varies
by substantial amounts over the distance  between resonance and corotation that varies
inversely with $K_y$.
For smooth solutions, the terms in equation (\ref{c0}) involving $d_i$ can be ignored in for large $K_y$. 
The smoothness condition on $Re(p)$ that determines the torque (see equation (\ref{Ti})) is expected hold as a consequence of the $x$ boundary condition (\ref{xbc}).

The series solution for $X <0$ follows similarly, however, there is a jump in $P'$ across the resonance.
The imaginary part of $P'$ undergoes a jump in value, as can be seen
by integrating equation (\ref{peq1}) in a small region about $X=0$ 
\begin{equation}
s_2(0) \int_{0-}^{0+} P''(X) dX =  s_2(0) P'(X) |_{0-}^{0+} = s_0(0) \int_C \frac{P(0)}{X} d X=   i \, \pi P(0) s_0(0),
\end{equation}     
where we have taken the integral along a counterclockwise contour $C$ that is a small half circle in the upper half-plane
about the $1/X$ pole \citep{MS87}.%. This procedure is equivalent to including the effects of a small dissipative term
%in the equations of motion \citep{MS87}.
 The expansion for $P$ that is valid for small positive and negative $X$ values can then be written as
\begin{equation}
P(X) = c_0 + (c_1 X + c_2 X^2)\log{|X|} + i c_3 \,H(-X)\, X + d_1 X + d_2 X^2,
\label{psing1}
\end{equation}
where $H$ is the Heaviside step function and 
\begin{eqnarray}
c_3 &=& - \frac{ \pi P(0) s_0(0)}{s_2(0)} \\
       & \simeq & \frac{\pi  s_0(0)}{s_2(0)},
\label{c3}
\end{eqnarray}
where we have used the fact that $P(0) = c_0 \simeq -1$ for large $K_y$,
as seen in equation (\ref{c0}), in obtaining the last equation.
It can be shown that $c_i$ are real for large $K_y$. 

Having determined the form of the pressure near the resonance, we can
determine leading singular behavior of the velocities from equations (\ref{eq:u}) - (\ref{eq:s}).
As follows from equation (\ref{psing1}), near the buoyancy resonance $X=0$,
\begin{equation}
P'(X) \sim c_1 (1+ \log{|X|}) + i c_3 H(-X) + d_1. 
\label{psing2}
\end{equation}
To lowest order in $K_y^{-1}$ for a Keplerian disk, quantity $c_1$ is given
by
\begin{eqnarray}
c_1 %&=& \frac{N h (g - 
    %h N^2) ( \kappa^2-N^2)}{
 %4 A g^2 K_y} \\
    &=& -\frac{\sqrt{\gamma-1}}{3 \gamma^{5/2} K_y}.
 \label{c1}
 \end{eqnarray}

We find that $u \sim i k_1 P'(X)$, $v \sim k_2 P'(X)$, $w \sim k_3 \delta(X)$ near $X=0$ for real constants $k_i$
 and Dirac delta
function $\delta(X)$.
Consequently, $u$  and $v$ are logarithmically singular near the resonance.
The equation of mass conservation (\ref{eq:rho}) implies that $Im(\rho) \sim k_4 u'(X) + k_5 w(X) \sim k_6 \delta(X)$.
This out of phase mass density contribution then leads to the local resonant torque density.
%Another contribution comes from the $\partial_z w$  

\section{TORQUE MODIFICATION DUE TO MIDPLANE BOUNDARY CONDITION FOR CONSTANT VERTICAL GRAVITY  \label{zbc}}

The torque derivation that led to equation (\ref{T1}) ignored the effects of the boundary conditions
at the disk midplane discussed in Section \ref{bc}.
We consider here the effects of the midplane boundary condition that we take as $w(x, 0)=0$, equation (\ref{zbc}).  
For $X>0$, we determine homogenous solutions  for the pressure perturbation  of the form
\begin{equation}
p_h(X,Z) = [1 + (c_1 X + c_2 X^2)\log{(X)} + d_1 X+ d_2 X^2 ] \exp{(-\lambda Z)},
\label{phom}
\end{equation}
where $\lambda$ is a dimensionless constant that is to be determined.
Solutions to the linearized pressure perturbation equation can be obtained as in
the inhomogeneous case of Appendix A. In the homogeneous case, $c_0$ does not appear
because the solution has an arbitrary scale factor. 
Instead, the expanded pressure equations are solved for $\lambda$, $c_1$, and $c_2$.
In the limit of large $K_y$,
these equations imply that
\begin{equation}
\lambda = \frac{2 k_y h \sqrt{a}}{\sqrt{3} (\kappa^2-N^2)},
\label{lam}
\end{equation}
where
\begin{equation}
a=-A \left[ A (N^2+\kappa^2) + \sqrt{A^2(N^2+\kappa^2)^2+6 N^2 (\kappa^2-N^2)(\kappa^2-8A\Omega-N^2)} \right].
\end{equation}
For a Keplerian disk with $\gamma=5/3$, the above evaluates to $\lambda \simeq 2.5 k_y h$.

Near the resonance, the vertical velocity and density perturbation are simply related
by
\begin{equation}
\rho(X,Z) = \mp \frac{i \rho_0 N}{g} w(X,Z).
\end{equation}
Consequently, the requirement that $w=0$ at the disk midplane
implies that $\rho$ and therefore the torque density $d T/dz$
also vanish there.
The equation for the  density perturbation variation in $Z$ and thus $dT/dz$ are modified 
by an additional term that is a homogeneous
solution with $Z$ variation $\exp{(-\lambda |Z|)}$ to account for this boundary condition.
The resonant torque distribution in $z$ is then given by
\begin{equation}
\frac{d T}{dz} =   \pm \frac{ \pi^2}{4} r^2 \rho_{00} \Phi^2 \frac{N^3}{ A  \, g^2}  \left[  \exp{(-|z|/h)} - \exp{(-\lambda |z|/h)} \right]
\label{dTdz2}
\end{equation}
with $\lambda$ given by equation (\ref{lam}).

\section{SHEARING BOX CALCULATION WITH VARIABLE GRAVITY \label{SBv}}

Equations (\ref{eq:u}) - (\ref{eq:s}) can be solved by applying a  Fourier method to the $x$ coordinate
through the use of sheared coordinates as described in \cite{GT78, GT80}. We added a viscous
term $\nu_{w} \partial_z^2 w$ to the right-hand side of equation (\ref{eq:w}) to provide greater
stability for the numerical scheme. The $z$ coordinate is not transformed.
%We consider a sector in which the $x \times z$ domain is $(0, x_{\rm o}) \times (0, z_{\rm u})$. 

The Fourier transformations are
\begin{equation}
\hat{q}(\tau, z) = \frac{1}{2 \pi} \int_{-\infty}^\infty q(x, z) \exp{(- i k_{y} \tau x)} d x,
\label{ft}
\end{equation}
where $q$ is $\Phi$, $u$, $v$, $w$, $\rho$, or $p$ and 
\begin{eqnarray}
\tau &=& k_x/k_y - 2 A t.
\end{eqnarray}
The inverse transformations are given by
\begin{equation}
q(x, z) = k_y \int_{\tau_{\rm s}}^{\tau_{\rm f}}  \hat{q}(\tau, z) \exp{( i k_{y} \tau x)} d \tau.
\label{qi}
\end{equation}
The limits of integration should ideally extend to  
$\tau_{\rm s} = -\infty$,  and $\tau_{\rm f} = \infty$. In the numerical scheme,
they are of course limited.

The transformed dynamical equations  (\ref{eq:u}) - (\ref{eq:s}) are respectively
\begin{eqnarray}
\partial_{\tau} \hat{u} - 2 \tilde{\Omega} \hat{v} & = &- i k \tau \left( \frac{ \hat{p}}{\rho_0} + \hat{\Phi} \right), \label{eq:u2v}\\
\partial_{\tau} \hat{v}  +2 \tilde{B} \hat{u} &= & -i k \left( \frac{ \hat{p}}{\rho_0} + \hat{\Phi} \right), \label{eq:v2v}\\
-2 A \partial_{\tau} \hat{w}  &=& -g\frac{\hat{\rho}}{\rho_{0}}
-\frac{\partial_{z} \hat{p}}{\rho_{0}} -\partial_z \hat{\Phi} + \nu_{w} \partial_z^2 \hat{w}, \label{eq:w2v}\\
\partial_{\tau} \hat{\rho} - \frac{1}{2 A} \hat{w}\partial_{z} \rho_{0}
 &=& - \rho_{0} \left(i k \tau \hat{u} +  i k \hat{v} - \frac{1}{2 A}\partial_{z} \hat{w} \right), \label{eq:rho2v}\\
\partial_{\tau} \left(\frac{\hat{p}}{c^2}-\gamma \hat{\rho} \right) &=&
\frac{ \rho_0 \hat{w} \, (\gamma-1)}{2 A } \frac{z}{h^2}  \label{eq:s2v},
\end{eqnarray}
where
\begin{eqnarray}
\tilde{\Omega} &=& -\frac{\Omega}{2 A},\\
\tilde{B} &=& -\frac{B}{2 A},\\
k &=& -\frac{k_y}{2 A}.\\
\end{eqnarray}

These are partial differential equations with a time-like coordinate $\tau$ and
a spatial coordinate $z$.
Guided by Section \ref{bc}, we apply the following boundary conditions at the disk midplane and upper boundary $z_{\rm u}$
\begin{eqnarray}
\hat{w}(\tau,z=0) &=& 0,\\
\hat{p}(\tau,z=z_{\rm u}) &=& 0,\\
\hat{w}(\tau,z=z_{\rm u}) &=& 0.
\end{eqnarray}

We consider the simple potential in which $\Phi(x,z)$ is real constant (equation (\ref{eq:Phic})).
\begin{eqnarray}
\hat{\Phi}(\tau, z) &=& \frac{1}{2 \pi} \int \Phi \exp{(- i k_{y} \tau x)} d x \\
            &=& \frac{\Phi}{k_y} \, \delta(\tau).
\end{eqnarray}
Since the potential turns on at $\tau=0$, we assume that
for $\tau <0$, all perturbed quantities are zero, e.g., $\hat{q}=0.$ 
We integrate the above equations over a very short interval centered at $\tau=0$ and obtain
\begin{eqnarray}
\hat{u}(0+,z) &=& 0 \label{eq:u3}, \\
\hat{v}(0+,z) &=&  \frac{i \,\Phi}{2 A}, \\
\hat{w}(0+,z) &=& 0, \\
\hat{\rho}(0+,z) &=& 0, \\
\hat{p}(0+,z) &=& 0. \label{eq:p3}
\end{eqnarray}

We then we integrate equations (\ref{eq:u2v}) - (\ref{eq:s2v}) in $\tau$ starting with $\tau=0$, subject to the initial conditions
(\ref{eq:u3}) - (\ref{eq:p3}) with $\hat{\Phi}=0$ (homogeneous equations). 

We apply the following boundary conditions
\begin{eqnarray}
\hat{u}(0,z) &=& 0,  \label{eq:ubc}\\
\hat{v}(0,z) &=&  \frac{i \,\Phi}{2 A}, \\
\hat{w}(0,z) &=& 0, \\
\hat{\rho}(0,z) &=& 0, \\
\hat{p}(0,z) &=& 0,\\
\hat{w}(\tau,0) &=& 0,\\
\hat{w}(\tau,z_{\rm u}) &=& 0,\\ 
\hat{p}(\tau,z_{\rm u}) &=& 0.  \label{eq:pbc}
\end{eqnarray}

We need to obtain the  density perturbation $\rho(x,z)$ in order to
determine the torque distribution. The inverse Fourier transform (\ref{qi}) 
contains artificial, small amplitude, rapid oscillations of density in $x$, due to the
finite duration of  $\hat{\rho}$ to $\tau=\tau_{\rm f}$ and the lack of its periodicity
over this interval.
To remove these unwanted oscillations, we multiply $\hat{\rho}(\tau,z)$ by a window (or tapering)
function that drops to nearly zero at $\tau_{\rm f}$ (e.g., \cite{P92}).
%Before taking its inverse Fourier transform (\ref{qi}), we apply a window or tapering function
%to remove artificial, small amplitude,  rapid oscillations of density in $x$ 
%that drops to nearly zero at $\tau_{\rm f}$ (e.g., \cite{H78}).  
%These oscillations come about because we are determining the transform
%at x-values that are more closely space than the  
%because the Fourier method implicitly assumes
%that the solutions $\hat{\rho}(\tau,z)$ are periodic in $\tau$ over time interval $(0, \tau_{\rm f})$.
%This phenomenon is known as spectral leakage in signal processing
%(e.g., \cite{H78}).  
%To removed these unwanted oscillations, we multiply $\hat{\rho}(\tau,z)$ by a window (or tapering)
%function that drops to nearly zero at $\tau_{\rm f}$ (e.g., \cite{H78})..
The window function was taken to be a Gaussian that is then applied to the density
as
\begin{equation}
\rho(x, z) = k_y \int_{0}^{\tau_{\rm f}}  \hat{\rho}(\tau, z)  \exp{(- 4 (\tau/\tau_{\rm f})^2)} \exp{( i k_{y} \tau x)} d \tau.
\label{qi1}
\end{equation}
This procedure eliminated the unwanted oscillations.
It gives similar results to those obtained by applying sufficiently strong viscous damping
in the radial and azimuthal dynamical
 equations (\ref{eq:u2v}) and (\ref{eq:v2v}), respectively. 
 Fast Fourier transforms were applied in Mathematica to evaluate the integral in equation (\ref{qi1}). 

In Fourier space, the outer buoyancy torque produces low frequency oscillations
in $\tau$ corresponding to a density disturbance at small $0 < x < h$.  
Since an outer Lindblad resonance wave occurs further out in $x$, it produces higher
frequency disturbances in $\tau$. Determining the properties of the buoyancy resonance
then involves separating the low from high frequency signals.

\newpage

\clearpage
\begin{thebibliography}


\bibitem[Bate et al.(2002)]{Bate02} Bate, M.~R., Ogilvie, 
G.~I., Lubow, S.~H., \& Pringle, J.~E.\ 2002, \mnras, 332, 575 

\bibitem[Goldreich 
\& Nicholson(1989)]{GN89} Goldreich, P., \& Nicholson, P.~D.\ 1989, \apj, 342, 1079 

\bibitem[Goldreich 
\& Tremaine(1978)]{GT78} Goldreich, P., \& Tremaine, S.\ 1978, \apj, 222, 850 

\bibitem[Goldreich \& Tremaine (1980)]{GT80} Goldreich, P. \& Tremaine, S. 1980, \apj, 241, 425

%\bibitem[Harris (1978)]{H78} Harris, F.~J. 1978, Proc. of the IEEE, 66, 1, p. 51
\bibitem[Goodman(1993)]{G93} Goodman, J.\ 1993, \apj, 406, 
596

\bibitem[Kley 
\& Nelson(2012)]{KN12} Kley, W., \& Nelson, R.~P.\ 2012, \araa, 50, 211 

\bibitem[Lin 
\& Papaloizou(1986)]{LP86} Lin, D.~N.~C., \& Papaloizou, J.\ 1986, \apj, 309, 846

\bibitem[Lubow \& Pringle (1993)]{LP93} Lubow, S.H., Pringle, J.E., 1993, ApJ, 409, 360

\bibitem[Lubow 
\& Ogilvie(1998)]{LO98} Lubow, S.~H., \& Ogilvie, G.~I.\ 1998, \apj, 504, 983 


\bibitem[Lubow 
\& Ida(2011)]{LI11} Lubow, S.~H., \& Ida, S.\ 2011, Exoplanets, edited by S.~Seager.~ Tucson, AZ: University of Arizona Press, 2011, 526 pp.~ ISBN 978-0-8165-2945-2., p.347-371 

\bibitem[Meyer-Vernet \& Sicardy (1987)]{MS87} Meyer-Vernet, N.\& Sicardy, B., 1987, Icarus, 69, 157

\bibitem[Ogilvie(1998)]{O98} Ogilvie, G.~I.\ 1998, \mnras, 
297, 291 

\bibitem[Press et al.(1992)]{P92} Press, W.~H., Teukolsky, 
S.~A., Vetterling, W.~T., 
\& Flannery, B.~P.\ 1992, Cambridge: University Press, |c1992, 2nd ed.,  

\bibitem[Vishniac et al.(1990)]{VD90} Vishniac, E.~T., Jin, 
L., \& Diamond, P.\ 1990, \apj, 365, 648 

\bibitem[Ward (1997)]{W97} Ward, W.~R.\ 1997, \apjl, 482, L211 

\bibitem[Zahn (1975)]{Z75} Zahn, J.-P.\ 1975, \aap, 41, 329 

\bibitem[Zhu et al  (2012)]{Zhu12} Zhu, Z., Stone, J.~M.,  \& Rafikov, R.~R.\ 2012, \apjl, 758, L42 

\end{thebibliography}
\end{document}